\documentclass[11 pt, oneside]{article}   	
\usepackage[margin=0.9 in]{geometry}                		
\usepackage{graphicx}				
\usepackage{amssymb}

\usepackage{algorithm}
\usepackage[noend]{algpseudocode}

\usepackage{cite}
\usepackage{amssymb,amsfonts}
\usepackage{algpseudocode}
\usepackage{multicol}
\usepackage{graphicx}
\usepackage{textcomp}
\usepackage{xcolor}
\usepackage{flushend}
\usepackage{amsmath}
\usepackage{amssymb}
\usepackage{amsfonts}
\usepackage{float}
\usepackage{graphicx}
\usepackage{fancyref}
\usepackage{booktabs}
\usepackage{subfigure}
\usepackage[affil-it]{authblk}

\graphicspath{ {./Modified_ERA/} }

\usepackage{bm}
\usepackage{tikz}
\usetikzlibrary{shapes.geometric, arrows}
\tikzstyle{startstop} = [rectangle, rounded corners, minimum width=3cm, minimum height=1cm,text centered, draw=black, fill=white!30]
\tikzstyle{io} = [trapezium, trapezium left angle=70, trapezium right angle=110, minimum width=3cm, minimum height=1cm, text centered, draw=black, fill=white!30]
\tikzstyle{process} = [rectangle, minimum width=3cm, minimum height=1cm, text centered, draw=black, fill=white!30]
\tikzstyle{decision} = [diamond, minimum width=3cm, minimum height=1cm, text centered, draw=black, fill=white!30]
\tikzstyle{arrow} = [thick,->,>=stealth]

\title{\vspace{-0 cm} \textbf{Towards an improved Eigensystem Realization Algorithm for low-error guarantees}}
\author{Mohammad N. Murshed
\thanks{Lecturer} }
\author{Moajjem Hossain Chowdhury
\thanks{Research Assistant} }
\author{Md. Nazmul Islam Shuzan
\thanks{Research Assistant} }
\author{M. Monir Uddin
\thanks{Associate Professor} }
 \affil{Department of Mathematics and Physics\\North South University\\ Dhaka, Bangladesh}

\begin{document}
\maketitle

\section*{\centering Abstract}
Eigensystem Realization Algorithm (ERA) is a tool that can produce a reduced order model (ROM) from just input-output data of a given system. ERA creates the ROM while keeping the number of internal states to a minimum level. This was first implemented by Juang and Pappa (1984) to analyze the vibration of aerospace structures from impulse response. We reviewed ERA and tested it on single input single output (SISO) system as well as on multiple input single output (MISO) system. ERA prediction agreed with the actual data. Unlike other model reduction techniques (Balanced truncation, balanced proper orthogonal decomposition), ERA works just as fine without the need of the adjoint system, that makes ERA a promising, completely data-driven, thrifty model reduction method. In this work, we propose a modified Eigensystem Realization Algorithm that relies upon an optimally chosen time resolution for the output used and also checks for good performance through frequency analysis. Four examples are discussed: the first two confirm the model generating ability and the last two illustrate its capability to produce a low-dimensional model (for a large scale system) that is much more accurate than the one produced by the traditional ERA.

\section*{\centering Introduction}
We consider the discrete linear system,
\begin{equation}
\textbf{x}_{i+1} = \textbf{A}\textbf{x}_{i} + \textbf{B} \textbf{u}_{i}
\label{S1}
\end{equation}
\begin{equation}
\textbf{y}_{i} = \textbf{C}\textbf{x}_{i} + \textbf{D}\textbf{u}_{i}
\label{S2}
\end{equation}
where \(\textbf{x} \in \mathbb{R}^{n}\) contains the internal states, \(\textbf{u} \in \mathbb{R}^{p}\) is the input, \(\textbf{y} \in \mathbb{R}^{q}\) refers to the output, and \(i\) is the time index. A set of inputs can cause a system to react in a particular manner via the internal states to result in the output. \(\textbf{A} \in \mathbb{R}^{n \times n}\), \(\textbf{B} \in \mathbb{R}^{n \times p}\), \(\textbf{C} \in \mathbb{R}^{q \times n}\) and \(\textbf{D} \in \mathbb{R}^{n \times p}\) are called the realizations.\\ \\
The number of equations in the system appears to be large for most of the common, complex problems around us. An example is the Navier Stokes equation where the system is large owing to the high number of spatial nodes in the domain. It is very difficult to extract knowledge from such large systems. Our focus in this work is on Eigensystem Realization Algorithm that makes a model from just impulse response data. We begin by reviewing some of the existing powerful model reduction techniques. \\ \\ 
 Model Order Reduction (MOR) is a way of reducing the complexity of models by means of projection. MOR aids in creating a low-dimensional version of the large scale system and enables a good enough understanding of the phenomenon in terms of fewer dominant states.This notion has been used to analyze random generated linear systems, flow past a flat plate, combustion and many other problems of interest. \\ \\
Proper Orthogonal Decomposition (POD) is a statistical method to derive a low rank version of a set of data, \cite{chatterjee2000introduction}. The idea can be tailored to obtain a reduced order model, but research has been in progress to find out better projections than the orthogonal ones. An interesting study of POD in the field of turbulence is available in \cite{sirovich1987turbulence}. Balanced truncation (Moore, 1981) and balanced proper orthogonal decomposition (BPOD) are two powerful model reduction methods successfully implemented on CFD problems and randomly generated systems \cite{willcox2002balanced}. Rowley et. al. used the idea of model reduction and extended it to solve non-linear complex compressible flows \cite{rowley2004model, rowley2005model}. The available techniques rely on the direct system and a transformed system also known as the adjoint system.\\ \\
Eigensystem Realization Algorithm, \cite{pappa1984galileo,juang1985eigensystem}, is the only model reduction tool that is based just on the direct system, hence, making it applicable on experimental data. It leverages just the input and output measurements to create a reduced order model for a given problem. The connection between ERA and BPOD is shown in \cite{ma2011reduced}. ERA has been tested on many occasions. It works well to make low-order models for unstable flows, \cite{flinois2016feedback}, which are then used to design controllers. Tangential interpolation based eigensystem realization algorithm (TERA), \cite{kramer2016tangential}, built on ideas of ERA to handle the huge amount of input-output data for multi-input multi-output (MIMO) system. TERA is applied on mass spring damper system and a cooling model for steel. A modified ERA, \cite{ma2011reduced}, is also developed and compared with the performance of balanced POD on the flow past a flat plate at a low Reynolds number.\\ \\
Since measured data may contain noise, the way the data gets separated into the signal and the noise has also been a subject under study. A noteworthy work can be found in \cite{li2011noise} that discusses the effect of noise, if any, on the modal parameters for a system. \\ \\
This paper is about the development of an improved version of Eigensystem Realization Algorithm that monitors and empirically determines the rank and the time resolution of the output measurement to produce a reduced order model that is much more reasonable than the one from conventional Eigensystem Realization Algorithm.
\begin{figure}
\centering
\begin{tikzpicture}[node distance=2cm]
\node (start)[startstop]{Input/Output Measurement};
\node (pro2a) [process, below of=start, yshift=-0.5cm] {Hankel (H) and time shifted Hankel Matrix (H') };
\node (dec1) [decision, below of=pro2a, yshift=-0.5cm] {SVD of H};
\node (out1) [io, below of=dec1,yshift=-0.5 cm] {\(A_{r},B_{r}, C_{r}\)};
\draw [arrow] (start) -- (pro2a);
\draw [arrow] (pro2a) --(dec1);
\draw [arrow] (dec1) --(out1);
\end{tikzpicture}\\
\caption{Eigensystem Realization Algorithm}
\label{fig:Arch}
\end{figure}
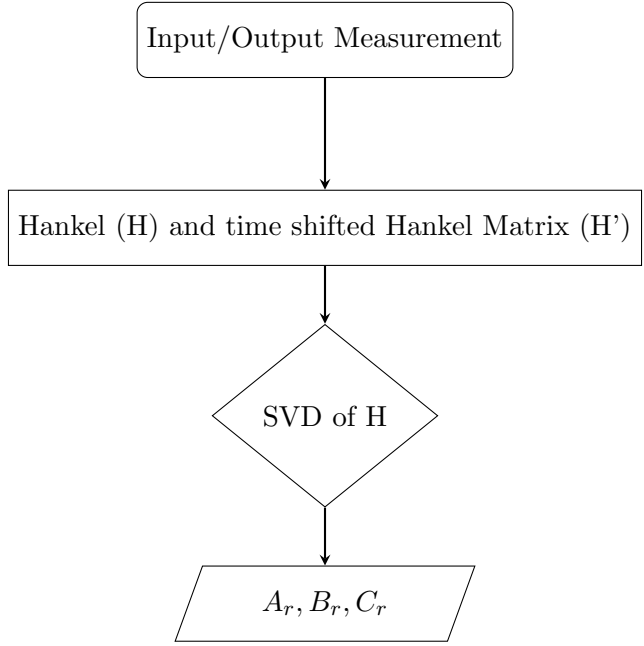

\section*{\centering Background}
In this section, we review the Eigensystem Realization Algorithm, its derivation and how it relates to Dynamic Mode Decomposition. These will be used to elaborate on the modified Eigensystem Realization Algorithm in the next section.

\subsection*{Eigensystem Realization Algorithm}
Eigensystem Realization Algorithm is a system identification method that was first used to create models for vibration in aerospace structures. This tool borrows from the idea of Ho's algorithm, \cite{zeiger1974approximate}, to find the realizations while keeping the number of internal states to a minimum that is to say that keeping the dimension of matrix \(A\) as low as possible. Completely data-driven, ERA uses only impulse response of the system i.e. just the inputs and the outputs, \cite{kutz2013data}. \\ \\
The discrete, linear time-invariant system in Eq.(\ref{S1}) and Eq.(\ref{S2}) can be excited by a pulse defined as
\[ u=\begin{cases} 

     1 & \text{\(k = 0\)}\\

      0 & \text{\(k > 0\)}. \\

   \end{cases}
\]
For \(x_{0} = 0\), the system reduces to \(x_{1}=Ax_{0} + B(1)\) to give \(x_{1}=B\). We can iterate through the system to get,
$$ x_{2} =Ax_{1} + B u_{1} = AB $$
$$ x_{3} =Ax_{2} + B u_{2} = A^{2}B $$
$$ x_{4} =Ax_{3} + B u_{3} = A^{3}B $$
and so on, while the outputs appear to be 
$$ y_{0} =Cx_{0} = 0 $$
$$ y_{1} =Cx_{1} = CB $$
$$ y_{2} =Cx_{2}  = CAB $$
$$ y_{3} =Cx_{3} = CA^{2}B $$
and so on. We observe that \(y_{k}=CA^{k-1}B\) which are also known as Markov parameters. Note that the dimension of Markov parameter is \(q \times p\). These are then used to construct the Hankel matrix and the time shifted Hankel matrix,
\[H=\begin{bmatrix}
    y_{1} & y_{2} & y_{3} &  y_{4} & ... & y_{m-s-1} \\
    y_{2} & y_{3} & y_{4} & y_{5} & ... & y_{m-s}\\
    \vdots & ... & ... & ... & \ddots & \vdots \\
    y_{s-1} & ... & ... & ... & ... & y_{m-2} \\
    \end{bmatrix}
    \]
\[H'=\begin{bmatrix}
    y_{2} & y_{3} &  y_{4} & y_{5} & y_{m-s-1} & y_{m-s} \\
    y_{3} & y_{4} & y_{5} & y_{6} & y_{m-s} & y_{m-s+1}\\
    \vdots & ... & ... & ... & \ddots & \vdots\\
    y_{s} & ... & ... & ... & y_{m-2} & y_{m-1} \\
    \end{bmatrix}.
    \]
    After performing the singular value decomposition of \(H=U\Sigma V^{*}\), the truncated version of \(U\),\(V\), \(\Sigma\) are computed as:
     $$ U_{r}=U(1:r,:) $$
     $$ V_{r}=V(1:r,:) $$
      $$ \Sigma_{r}=\Sigma(1:r,1:r),$$
      where r is the rank of \(H\). The reduced order model is then given by,
    $$ A_{r} = \Sigma^{-1/2} U_{r}^{*}H'V_{r}\Sigma^{-1/2} $$
    $$ B_{r} = \text{the first  \(p\) columns of} \ \Sigma_{r}^{1/2}V_{r}^{*}$$
    $$ C_{r} = \text{the first \(q\) rows of} \ U_{r}\Sigma_{r}^{1/2}.$$


There are a few important points about the notation used. The output measurement is a function of time and variable \(s\) controls the way we stack the time shifted output measurement. The Hankel matrices above refer to a single input and single output system (SISO). The dimensions would change as we deal with a different system e.g. MISO or MIMO. \\ \\
\textbf{Note on Hankel Singular Values.} In general, eigenvalues give hint on the system stability. But, Hankel singular values identify the highly energetic states that contribute the most to characterize the system. That means the states with low energy can be truncated to obtain an approximate model. The Hankel singular values are computed from the SVD of the product of the controllability and the observability Gramian. 
\subsection*{Connection to Dynamic Mode Decomposition}
Dynamic Mode Decomposition (DMD) is a strategy of creating a model from time series data, \cite{s2d}. It can be viewed as a special case of Koopman operator which is a way of representing a non-linear dynamical system as a infinite-dimensional linear system. DMD has numerous applications in various disciplines like fluid dynamics, neuroscience, epidemiology and many others. There are a lot of variants of DMD that are to be utilized based on the type of data. For instance, time delay coordinate based DMD is an option when the data is highly oscillatory \cite{9038561}.\\ \\ DMD aims to map the current states to the future states as,
$$ X_{i+1}= AX_{i}. $$
This equation from DMD is essentially Eq.(\ref{S1}) with no control. This implies that DMD is related to ERA from a dynamical systems point of view. 

%
%


\section*{\centering Modified Eigensystem Realization Algorithm}
We propose an improved version of Eigensystem Realization Algorithm that searches for the optimal number of temporal nodes, \(N_t\), used to express the output from the actual system and also the optimal rank, \(r\), used in ERA to attain a reduced order model that can predict the output with high accuracy. \\ \\
The idea behind this modified ERA is to run the conventional ERA multiple times so to identify the best possible number of temporal nodes and the rank that keeps the error,
\begin{equation}
\epsilon=|| \textbf{y}_{actual} - \textbf{y}_{ERA} ||_{2}
\label{epsil}
\end{equation}
as low as possible. \(N_{t}\) controls the time resolution: certain \(N_{t}\) values are optimal while others  can yield large error. The steps in the middle are the same as the ones in the traditional ERA. At the very end of this modified version, the \(A_{r}, B_{r}\), and \(C_{r}\) are computed based on the optimal rank. The routine is provided in Algorithm \ref{ALGO:1}. The specialty of this updated version of ERA is that it uses the appropriate time resolution and identifies the 'best' possible rank to keep the error to a minimum. 

\begin{algorithm}
\textbf{Pre-run ERA to identify \(N_t\)  and \(r\) that keep the error, (\ref{epsil}), as low as possible}\\
Utilize the output measurements, \(\textbf{y}_{actual}\), based on the optimal time resolution, \(T/N_{t}\)\\
Construct the Hankel matrix (\(H\)) and the time-shifted Hankel matrix (\(H' \))\\
Compute the SVD of the Hankel matrix,  \(H=U\Sigma V^{*}\)\\
Find the truncated \(U,\Sigma\), and \(V\) using the rank (\(r\)) from the pre-run\\
Calculate the reduced system matrices just as in traditional ERA\\
Generate the output from ERA, \(\textbf{y}_{ERA}\), via the reduced order model
\caption{\textbf{Modified Eigensystem Realization Algorithm}}
\label{ALGO:1}
\end{algorithm}
We also recommend that a frequency analysis be performed after this modified ERA scheme is enacted. Frequency analysis is often helpful for engineering purposes. A way to do it is by using \(tfestimate\) on MATlab. This function takes in the input and output to generate an approximate transfer function for a certain range of frequencies. Note that \(bodeplot\) on MATlab results in the magnitude and phase of the the system, but visual comparison is well done via \(tfestimate\).

\section*{\centering Numerical Results}
We have tested ERA on four different problems. The first two aim to stress on the model identification function of ERA and the last two prove the ability of ERA to work as a model reduction tool. The results are generated on a personal computer (HP Pavilion 14) with CORE i5 8th Gen processor 1.6-3.4 GHz and RAM of 8 GB via MATLAB version 2019b.  

%

\subsection*{Example 1. Pitch Model (SISO)}
The 3 D motion of an aircraft is governed by the pitch, plunge and surge models, \cite{brunton2014state}. Many state variables come into play. Velocity, density, temperature and pressure are a few of them. Computing all these state variables in a grid is not easy since there may be spatial nodes as many as \(10^{6}\). In aeronautics, engineers care a lot about what is called the lift coefficient per unit span,
$$ C_{L}= \frac{2L}{\rho U_{\infty}^{2} c}$$ 
where \(L\) is the lift force on the wing, \(\rho\) the air density and \(U_{\infty}\) free-stream velocity,and  \(c\) the chord. The angle of attack, \(\alpha\), is the angle between the airfoil chord and the flow direction. It can be thought of as some angular displacement which automatically makes \(\ddot{\alpha}\) the angular acceleration.
The pitch motion of an aircraft is the one that is observed with the nose moving up or down. We use the linearized pitch model from \cite{brunton2014state} that reads:
\[\frac{d}{dt}\begin{bmatrix}
    \textbf{x} & 0 & 0 \\
    \alpha & 0 & 0\\
    0 & 0 & 1 \\
    \end{bmatrix}=
   \begin{bmatrix}
    \textbf{A} & \textbf{0} & \textbf{B}_{\dot{\alpha}} \\
    \textbf{0} & 1 & 0\\
    \textbf{0} & 0 & 0 \\
    \end{bmatrix}\begin{bmatrix}
    \textbf{x}\\
    \alpha\\
    \dot{\alpha}\\
    \end{bmatrix}+ \begin{bmatrix}
    \textbf{0}\\
    0\\
    1\\
    \end{bmatrix} \ddot{\alpha}
    \]
    \[C_{L}=\begin{bmatrix}
    \textbf{C} & C_{\alpha} & C_{\dot{\alpha}} \\
    \end{bmatrix}\begin{bmatrix}
    \textbf{x}\\
    \alpha\\
    \dot{\alpha}\\
    \end{bmatrix} C_{\ddot{\alpha}}\ddot{\alpha}
    \]
where \textbf{x} is a vector containing the states. A non-dimensionalized version of time is used, \(\tau = t \frac{U_{\infty}}{c}\). \\
\begin{figure}
\centering
\subfigure[Angular Acceleration Variation]{\includegraphics[scale=0.4]{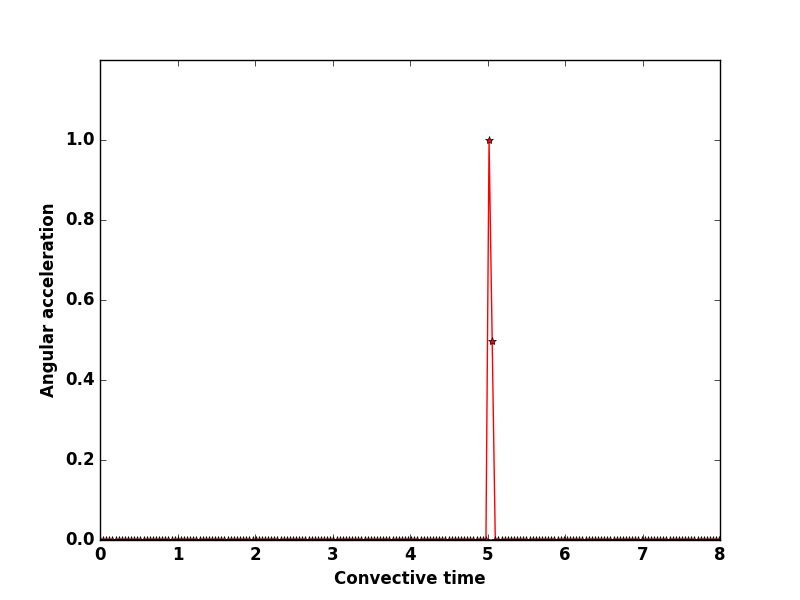}\label{alpha}}
\subfigure[Lift coefficient data and ERA prediction for pitch model]{\includegraphics[scale=0.4]{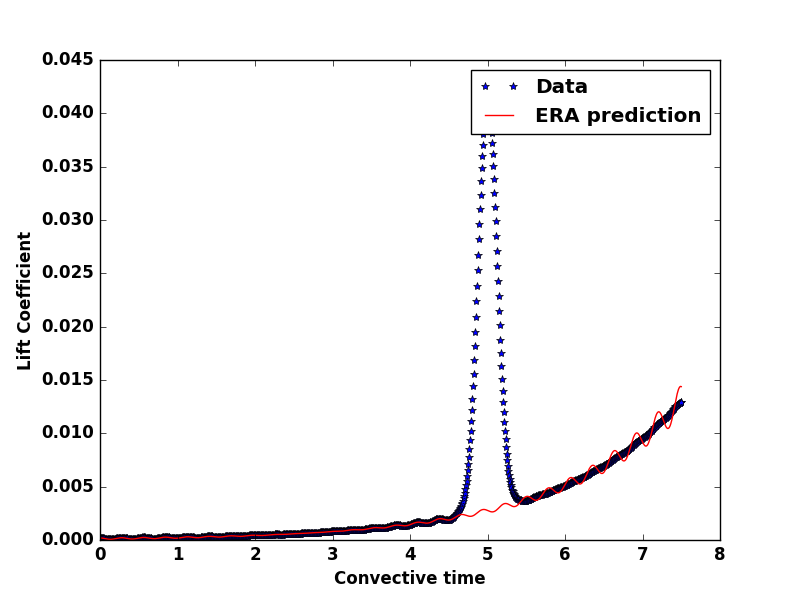}\label{cl}}
\caption{Pitch model Input and Response}
\end{figure}
\begin{figure}
\centering
\subfigure[Input: Unit Step function]{\includegraphics[scale=0.4]{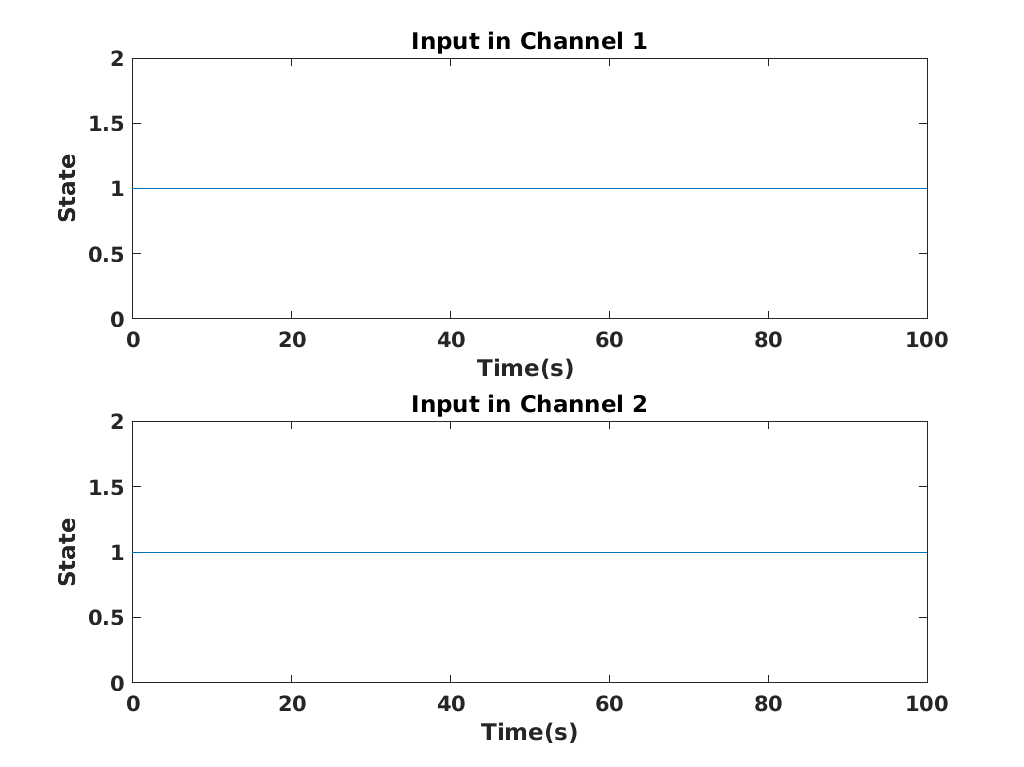}}
\subfigure[Input: Ramp function]{\includegraphics[scale=0.4]{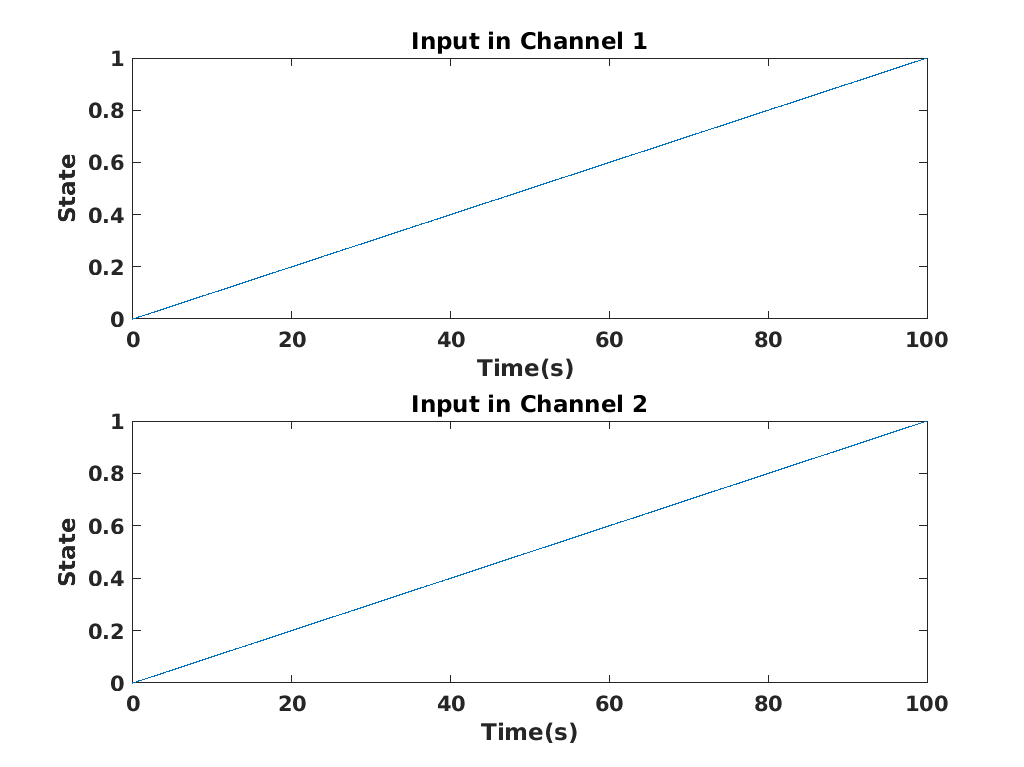}}
\subfigure[Input: Random numbers]{\includegraphics[scale=0.4]{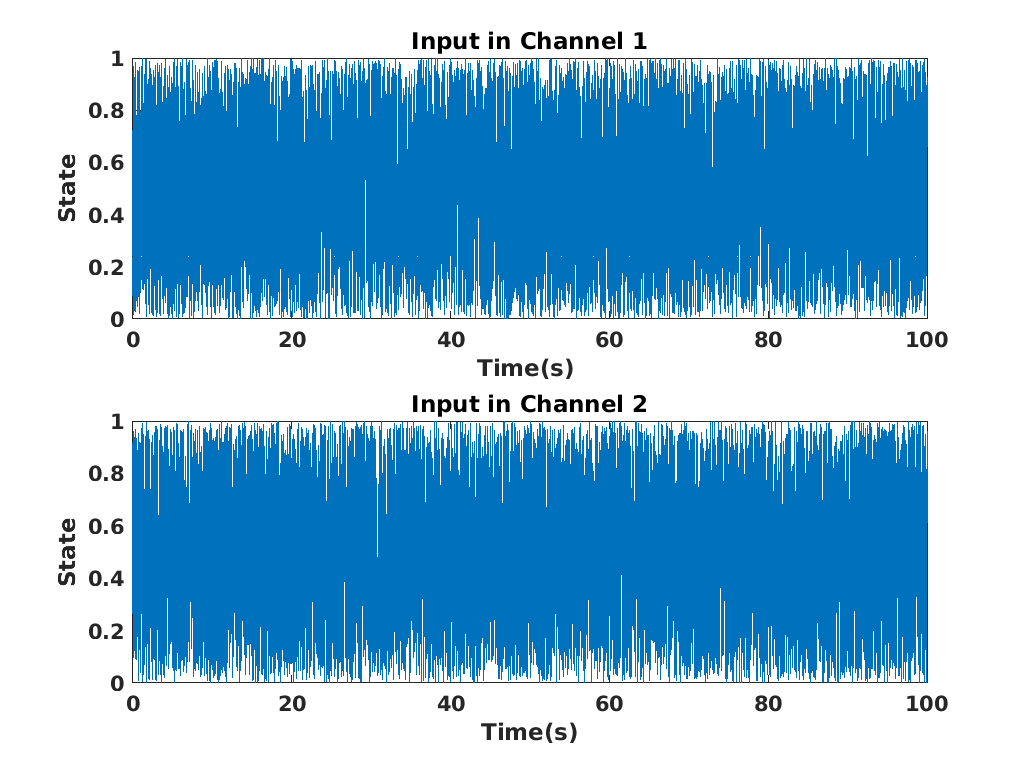}}
\caption{Arbitrary inputs used to check performance}
\label{Arb_inputs}
\end{figure}

\begin{figure}
\centering
\subfigure[Input:Dirac Delta]{\includegraphics[scale=0.45]{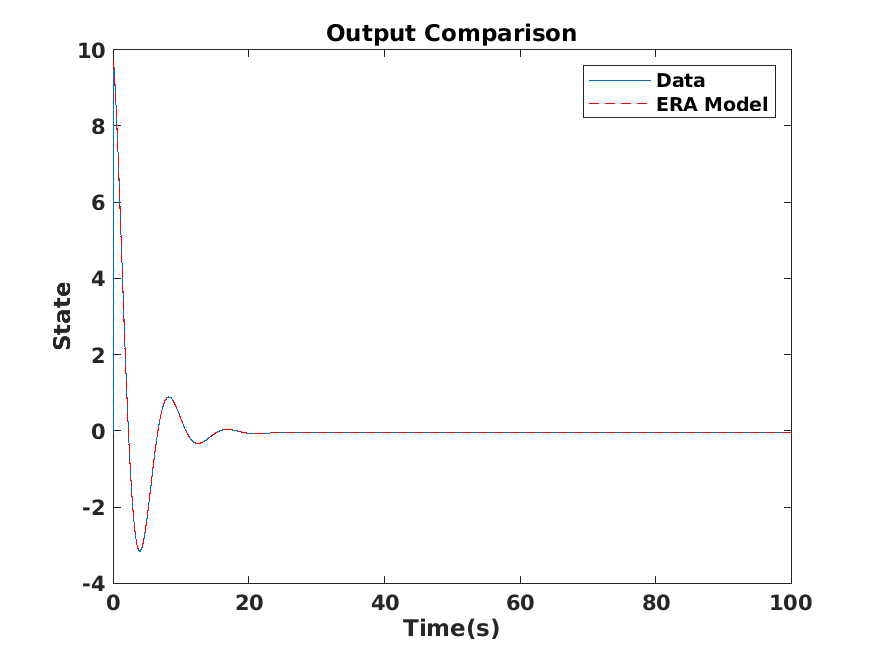}}
\subfigure[Input: Unit Step function]{\includegraphics[scale=0.45]{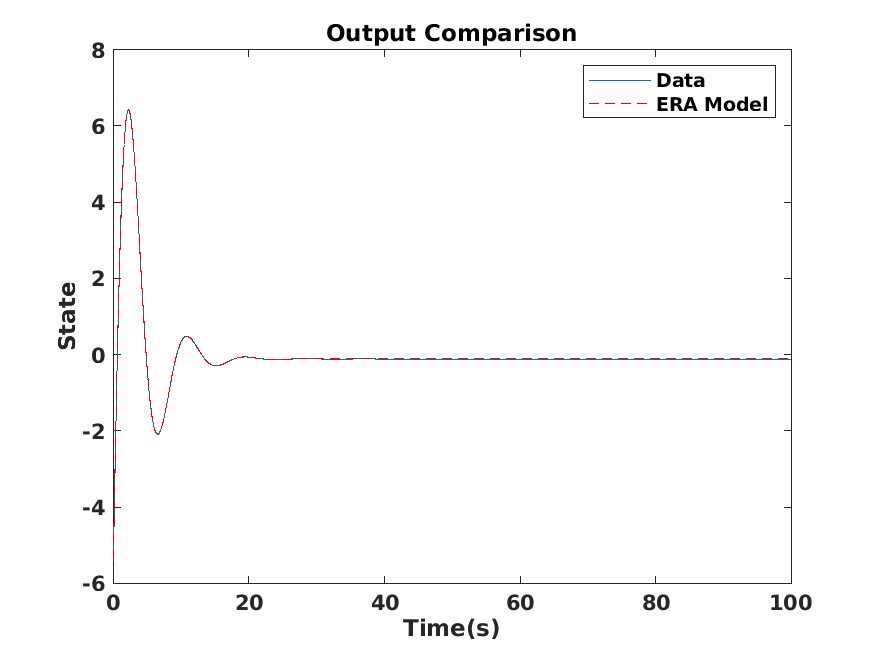}}
\subfigure[Input: Ramp function]{\includegraphics[scale=0.45]{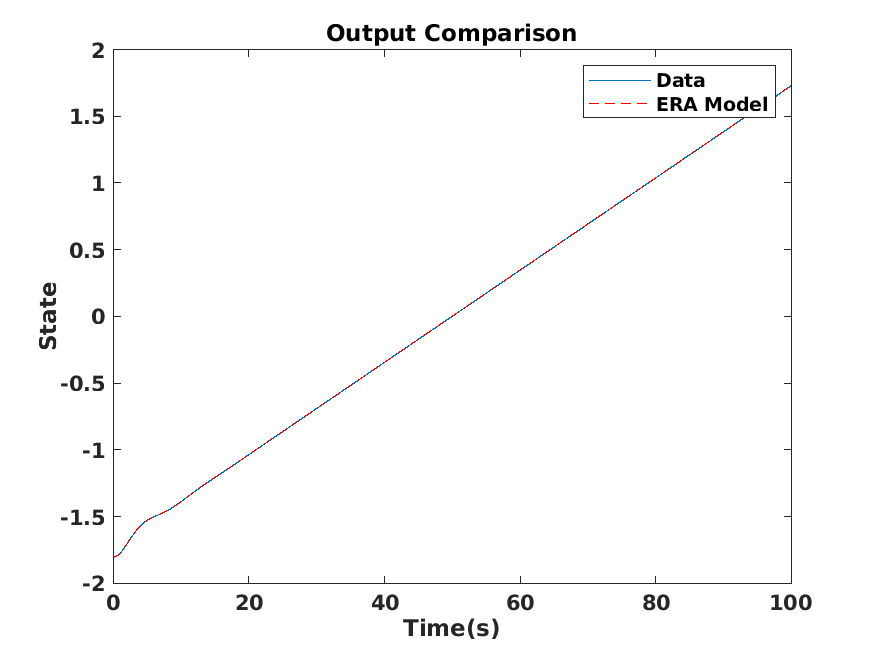}}
\subfigure[Input: Random numbers]{\includegraphics[scale=0.45]{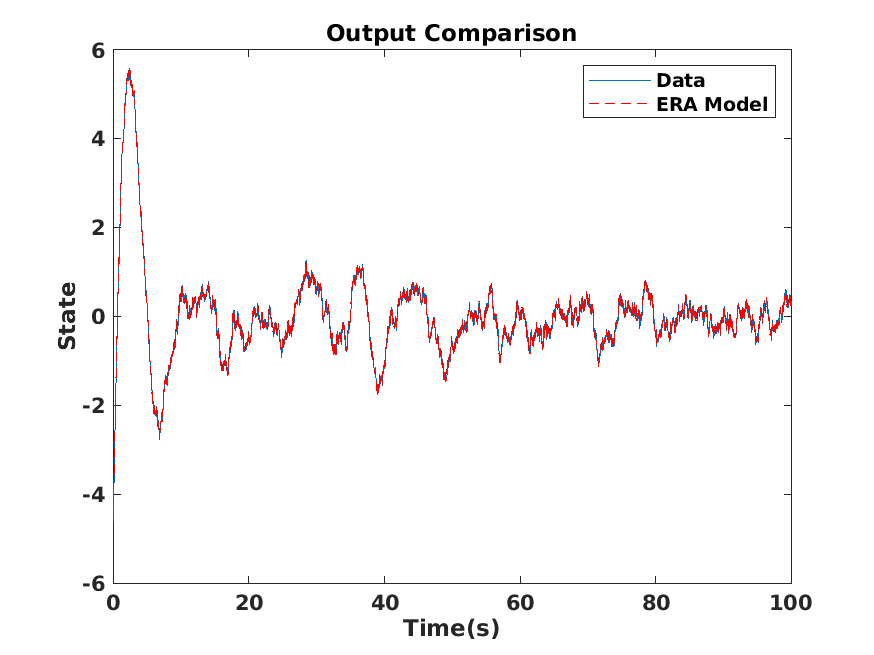}}
\caption{Response comparison between ERA and actual system for various inputs}
\label{RESP_Compare}
\end{figure}

The pitfall in applying ERA on a state space model is that we are bound to use an impulse response. Thus, it is worth correctly identifying the right input for the pitch motion of an aircraft. We examine the behavior of various state variables over time, available in \cite{brunton2014state}. The angle of attack and angular velocity vary in the form of a ramp and step function, respectively, whereas the angular acceleration, Figure \ref{alpha}, follows a dirac delta function which is the requirement of ERA. Hence, angular acceleration would be a suitable input. Note that we consider the input from \(\tau = 5\) upto \(\tau=7.5\). In this example, we use exponential functions to approximate the lift coefficient behavior, 
$$ C_{l}=\frac{1}{105}(\frac{1}{ (\sigma \sqrt(2 \pi))} e^{-0.5((\tau-5)/\sigma)^2} + 2.2^{0.68\tau-5.8}),$$
with \(\sigma\) set at \(\sqrt{0.015}\). This signal is fed into ERA to find the reduced order model for the pitch dynamics of an aircraft. Figure \ref{cl} illustrates that the data and ERA model agree for most of the time except where the lift coefficient has a sudden significant jump followed by a drop at around \(\tau=5\).

%

\subsection*{Example 2. Second Order State Space Model (MISO)}
We also test ERA on a multi input single output system and compare the reduced order model to the original model by feeding arbitrary inputs (step, ramp, unit and random). The MISO, in consideration, is defined as per the following matrices,
\[A=\begin{bmatrix}
    -0.5572 & -0.7814 \\
    0.7814 & 0\\
    \end{bmatrix};
    B=\begin{bmatrix}
    1 & -1 \\
    0 & 2\\
    \end{bmatrix}\]
    \[ C=\begin{bmatrix}
    1.9691 & 6.4493\\
    \end{bmatrix};D=\begin{bmatrix}
    0 & 0\\
    \end{bmatrix}.
    \]
    
%

The reduced order model extracted by ERA came out to be,
\[\hat{A}=\begin{bmatrix}
    0.9985 & 0.0137 \\
    -0.004 & 0.9959\\
    \end{bmatrix};
    \hat{B}=\begin{bmatrix}
    -1.7916 & -1.9913 \\
    -0.5389 & 2.1738\\
    \end{bmatrix}\]
    \[ \hat{C}=\begin{bmatrix}
    -2.0568 & 3.1113\\
    \end{bmatrix};\hat{D}=\begin{bmatrix}
    1.9691 & 10.9295\\
    \end{bmatrix}.
    \]

     After extracting the Markov parameters and creating a model, we tested the model with different inputs. The outputs, Figure \ref{RESP_Compare}, from these different inputs, Figure \ref{Arb_inputs}, are compared with outputs from the original model. The error between these outputs were minimal. Even when the input was a randomly generated signal, the model was able to capture the characteristics of the signal. The ERA output resembles the actual output despite the difference in the Markov parameters of the two systems. We can see that ERA found the parameters from the data.

    \subsection*{\textbf{Example 3.} Heat Diffusion Equation (Sparse Model)}
\begin{figure}
\centering
\subfigure[The sparsity pattern in state matrix, A]{\includegraphics[scale=0.5]{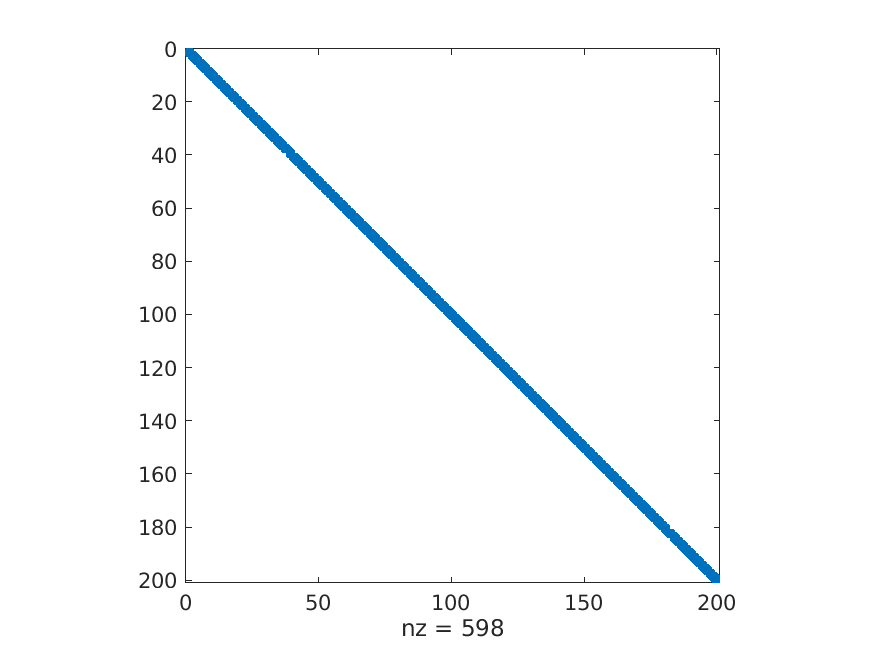}\label{sparseA}}
\subfigure[Comparison of the output from the actual system and ERA]{\includegraphics[scale=0.5]{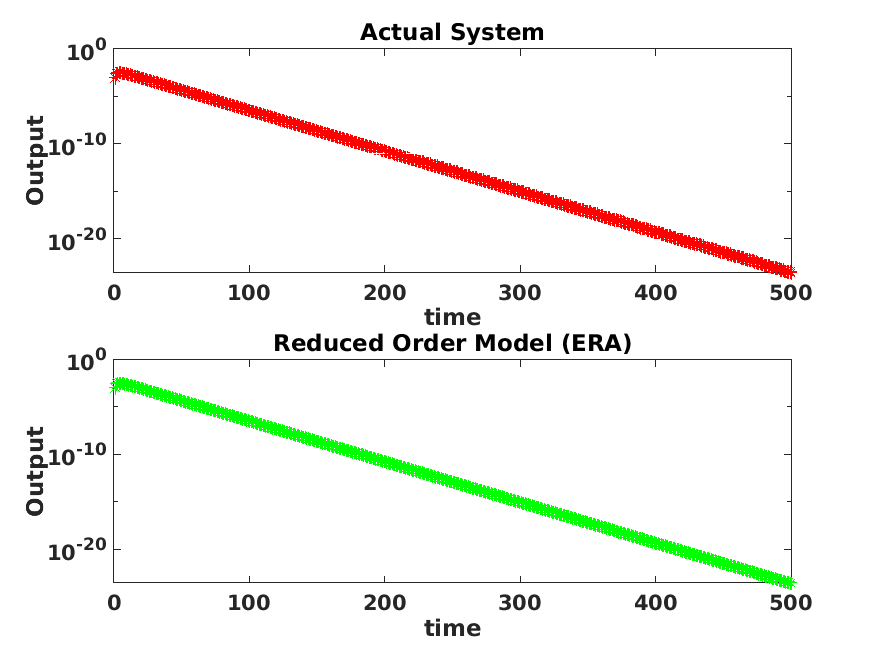}\label{asera}}
\subfigure[Uncertainty with time]{\includegraphics[scale=0.5]{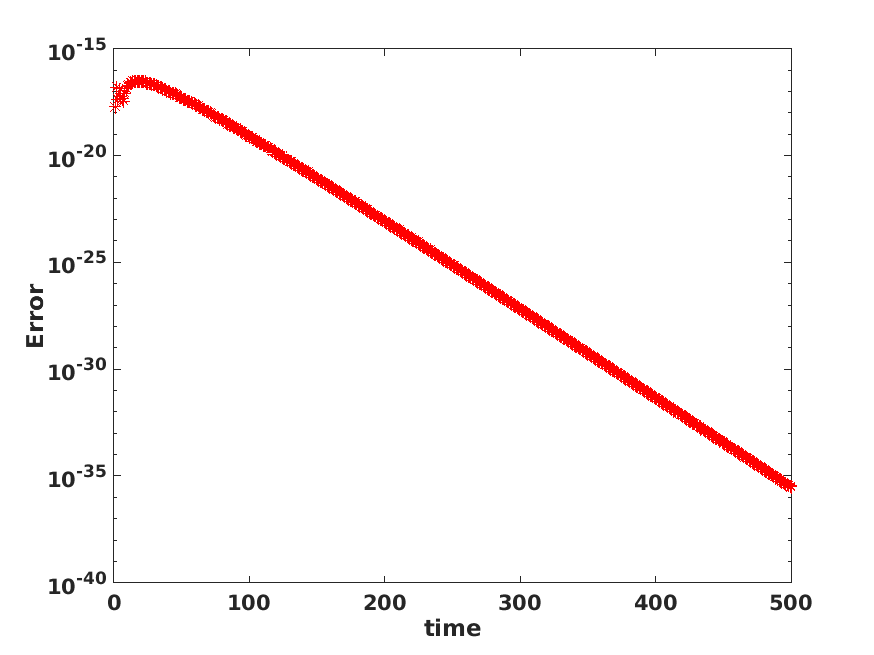}\label{err_heat}}
\subfigure[Error behavior with the number of temporal nodes]{\includegraphics[scale=0.5]{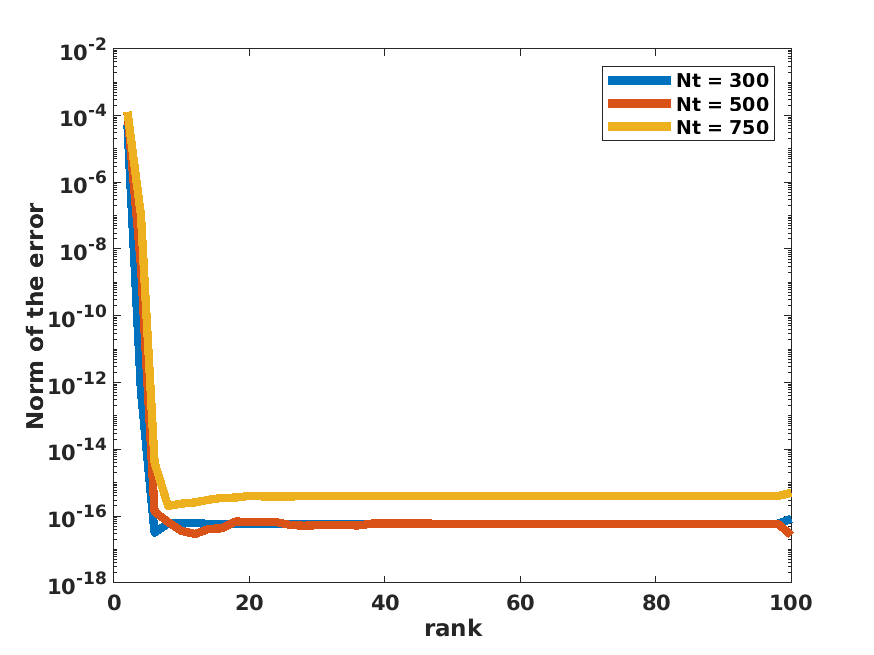}\label{Err_Nt_heat}}
\subfigure[Frequency analysis for different rank used in the ERA]{\includegraphics[scale=0.5]{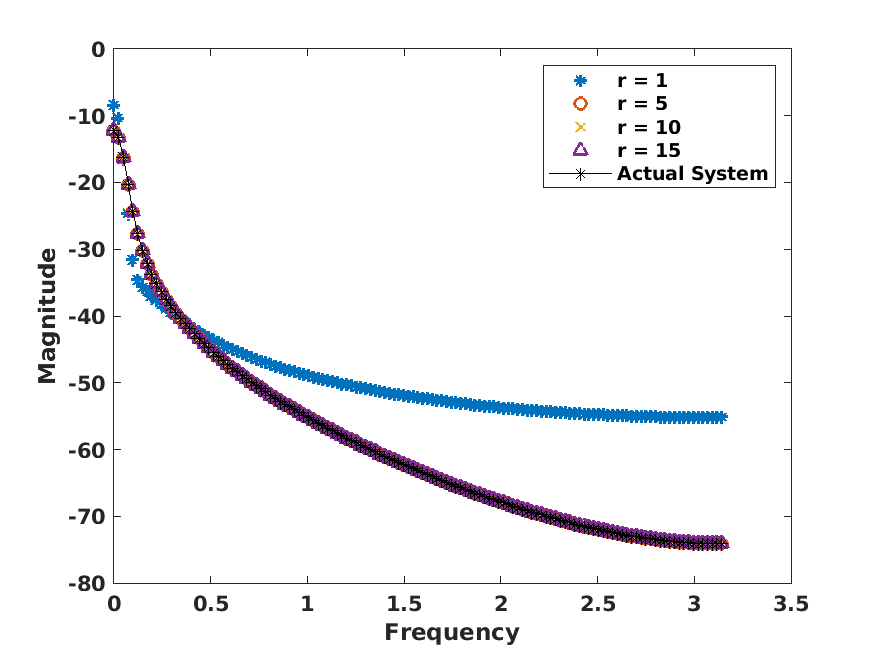}\label{heat_abso_freq}}
\caption{Heat-Diffusion model reduction}
\end{figure}

    In this example, the heat diffusion equation for a rod of unit length (1D) is defined as,
     $$ \frac{\partial T(x,t)}{\partial{t}} =\alpha \frac{\partial^{2} T}{\partial x^{2}} + u(x,t) $$
    $$ T(0,t) = T(1,t) = 0 $$
    $$ T(x,0) = 0, $$
    where \(x\) refers to space and \(t\) is the time. The equations above consist of the partial differential equation showing the evolution of the temperature, \(T(x,t)\), the boundary conditions, and the initial condition. The input is controlled by \(u(x,t)\). \\ \\
We utilized the system matrices available in \cite{chahlaoui2002collection} (the state matrix shown in Figure \ref{sparseA}) to compute the output for this problem. The output is then fed into ERA to identify a reduced order model. The crux is to observe the behavior of the norm of the error with the rank set in ERA as in Figure \ref{Err_Nt_heat}. This allows us to find the optimal number of temporal nodes to be used that maintains a low rank for the system. We have identified that \(N_{t}=300\) works fine for \(r=6\). Thus, the state matrix in the original system is reduced from \(200 \times 200 \) to \(6 \times 6\) via ERA. The output from ERA, produced by the \(lsim\) function, also agrees well with the output from the actual large system, Figure \ref{asera}. The error over time is also displayed in Figure \ref{err_heat}. The frequency analysis done by \(tfestimate\) shows that \(r\) greater 1 yields a model as good as the actual system, Figure \ref{heat_abso_freq}.

  \subsection*{\textbf{Example 4.} Atmospheric Storm Track (Dense Model)}
  The atmospheric storm track is a model from oceanography used to analyze the velocity of the airflow in the zonal (latitude wise) and meridional (longitude wise) setting. We can imagine this of a flow in a channel, the physical domain of which is defined as,
  $$ 0 <x<12 \pi$$ 
  $$ -0.5\pi <y<0.5\pi$$
   $$ 0<z<1. $$
   In the \(z\) axis, \(z = 0\) is the ground level and \(z = 1\) is the tropopause.\\
  The mean velocity is set to vary with the altitude,
  $$ U(z) = 0.2 + z, $$
and time is non-dimensionalized as \(T=\frac{L}{U_{0}}\) where \(L = 1000 \ km\) and \(U_{0}= 30 \ m/s\). The system is thought to have a uniform flow, but can be disturbed by a linear damping at the entrance and the exit of the track. The details of the dynamics can be found in \cite{chahlaoui2002collection}. The governing equation of the states is,
\begin{equation}
\frac{d\psi}{dt} =A\psi,
\end{equation}
where \(\psi\) is the velocity variable.\\ \\
\begin{figure}
\centering
\subfigure[Comparison of the output from the actual system and ERA]{\includegraphics[scale=0.5]{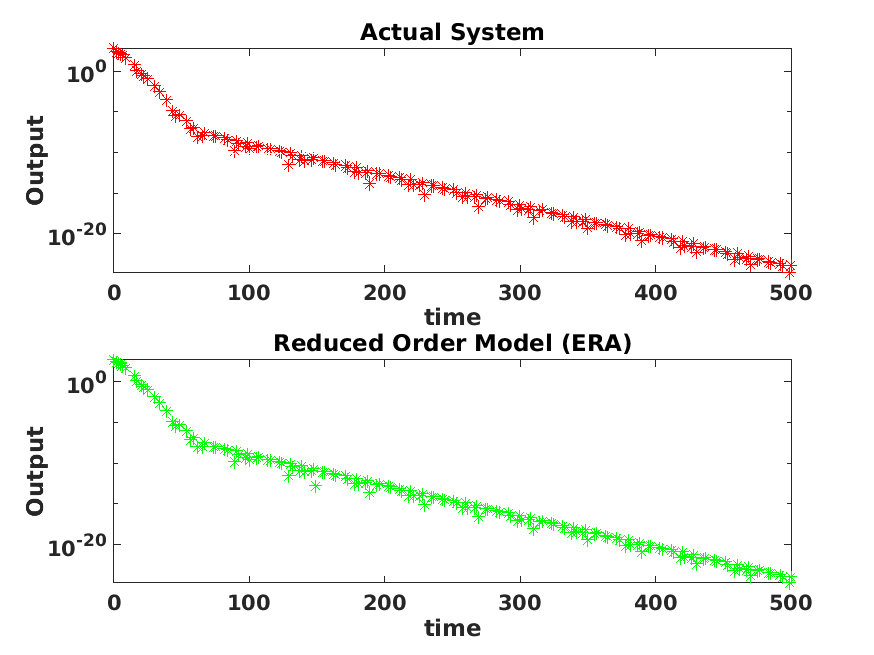}\label{comp_eady}}
\subfigure[Absolute value of the difference in the output from the actual system and ERA]{\includegraphics[scale=0.5]{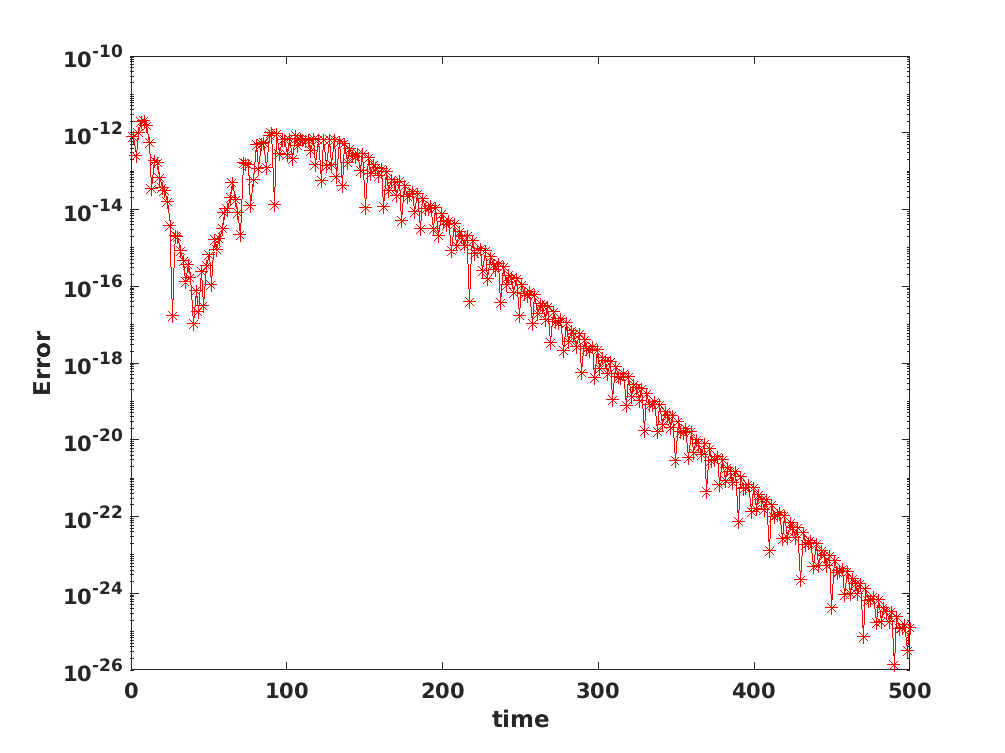}\label{Err_eady}}
\subfigure[Error behavior with the number of temporal nodes]{\includegraphics[scale=0.5]{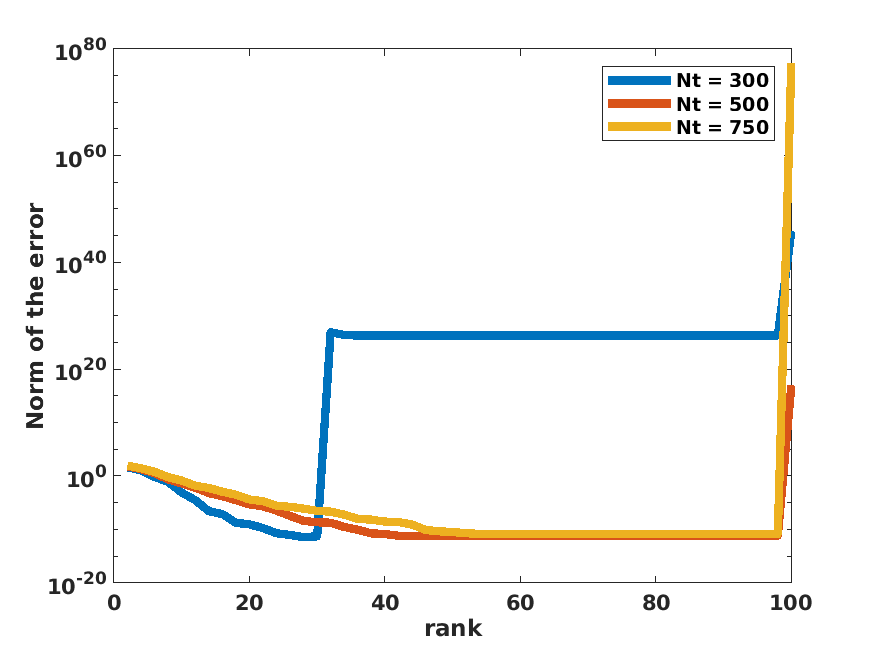}\label{Err_Nt}}
\subfigure[Frequency analysis for different rank used in the ERA]{\includegraphics[scale=0.5]{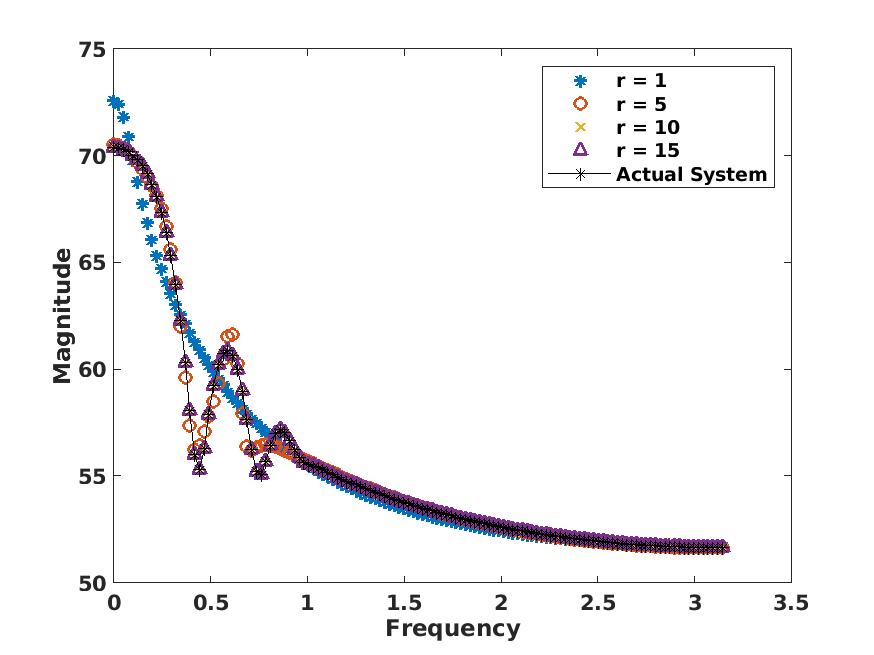}\label{eady_abso_freq}}
\caption{Atmospheric Storm Track model reduction}
\end{figure}
The output from the actual system is put into ERA. Figure \ref{Err_Nt} shows that \(r \approx 55\) for \(N_{t}=500\) and \(N_{t}= 750\) whereas use of 300 temporal nodes allows for \(r=30\). Setting the rank to 30 and number of temporal nodes to 300 results in ERA output that resembles the original output, \ref{comp_eady}. The norm of the difference between the approximate output and the actual output is also plotted in \ref{Err_eady}. We observe that the intractable original system of dimension \(598 \times 598\) gets reduced to a \(30 \times 30\) system by ERA.\\ \\
The transfer function estimate for several different rank values are plotted in Figure \ref{eady_abso_freq}. The ERA model with \(r=1\) is far away from the actual system, \(r=5\) and \(r=10\) show improvement and \(r=15\) enables reduced order modeling that is as efficient as the actual system. Thus, proper selection of the time resolution and rank along with a frequency analysis in the Eigensystem Realization Algorithm shows promise in building reduced order models with low error. 

\section*{\centering Conclusion and Future Work}
In this work, we delineated the steps in our proposed modified Eigensystem Realization Algorithm and implemented this method on four test problems. Modified ERA identifies the model in the first two examples and performs as a tool to reduce the order of the model in the third and fourth example. By model identification, we mean finding the state, input and output matrix and model order reduction refers to the minimization of the size of the state matrix, also known as the system matrix. The output predicted by ERA agree well with that from the original system. The first example is concerned about the pitching motion of an aircraft, and the second one a second order state space model. The third example is the heat-diffusion equation and the last one a model for the airflow velocity when a storm or a cyclone surges. Indeed, the third and the fourth numerical tests demonstrate that the rank should be carefully set at or above 5 to minimize error in the output predicted by ERA and also to get a transfer function estimate that is much close to that of the original system. \\ \\
We plan to work on a survey of all the model order reduction techniques and apply them on an array of synthetic and practical data and finally weigh the pros and cons of each technique. At the same time, our work would also be to establish any connection between model order reduction method and DMD, \cite{tu2012improved}. 

\section*{\centering Acknowledgement}
This project is partially funded by Office of Research, North South University. (Grant number: \textbf{CTRG-19/SEPS/06})

\bibliography{mor1}

\begin{thebibliography}{10}
\providecommand{\url}[1]{#1}
\csname url@samestyle\endcsname
\providecommand{\newblock}{\relax}
\providecommand{\bibinfo}[2]{#2}
\providecommand{\BIBentrySTDinterwordspacing}{\spaceskip=0pt\relax}
\providecommand{\BIBentryALTinterwordstretchfactor}{4}
\providecommand{\BIBentryALTinterwordspacing}{\spaceskip=\fontdimen2\font plus
\BIBentryALTinterwordstretchfactor\fontdimen3\font minus
  \fontdimen4\font\relax}
\providecommand{\BIBforeignlanguage}[2]{{%
\expandafter\ifx\csname l@#1\endcsname\relax
\typeout{** WARNING: IEEEtran.bst: No hyphenation pattern has been}%
\typeout{** loaded for the language `#1'. Using the pattern for}%
\typeout{** the default language instead.}%
\else
\language=\csname l@#1\endcsname
\fi
#2}}
\providecommand{\BIBdecl}{\relax}
\BIBdecl

\bibitem{chatterjee2000introduction}
A.~Chatterjee, ``An introduction to the proper orthogonal decomposition,''
  \emph{Current science}, pp. 808--817, 2000.

\bibitem{sirovich1987turbulence}
L.~Sirovich, ``Turbulence and the dynamics of coherent structures. i. coherent
  structures,'' \emph{Quarterly of applied mathematics}, vol.~45, no.~3, pp.
  561--571, 1987.

\bibitem{willcox2002balanced}
K.~Willcox and J.~Peraire, ``Balanced model reduction via the proper orthogonal
  decomposition,'' \emph{AIAA journal}, vol.~40, no.~11, pp. 2323--2330, 2002.

\bibitem{rowley2004model}
C.~W. Rowley, T.~Colonius, and R.~M. Murray, ``Model reduction for compressible
  flows using pod and galerkin projection,'' \emph{Physica D: Nonlinear
  Phenomena}, vol. 189, no. 1-2, pp. 115--129, 2004.

\bibitem{rowley2005model}
C.~W. Rowley, ``Model reduction for fluids, using balanced proper orthogonal
  decomposition,'' \emph{International Journal of Bifurcation and Chaos},
  vol.~15, no.~03, pp. 997--1013, 2005.

\bibitem{pappa1984galileo}
R.~Pappa and J.-N. Juang, ``Galileo spacecraft modal identification using an
  eigensystem realization algorithm,'' in \emph{25th Structures, Structural
  Dynamics and Materials Conference}, 1984, p. 1070.

\bibitem{juang1985eigensystem}
J.-N. Juang and R.~S. Pappa, ``An eigensystem realization algorithm for modal
  parameter identification and model reduction,'' \emph{Journal of guidance,
  control, and dynamics}, vol.~8, no.~5, pp. 620--627, 1985.

\bibitem{ma2011reduced}
Z.~Ma, S.~Ahuja, and C.~W. Rowley, ``Reduced-order models for control of fluids
  using the eigensystem realization algorithm,'' \emph{Theoretical and
  Computational Fluid Dynamics}, vol.~25, no. 1-4, pp. 233--247, 2011.

\bibitem{flinois2016feedback}
T.~L. Flinois and A.~S. Morgans, ``Feedback control of unstable flows: a direct
  modelling approach using the eigensystem realisation algorithm,''
  \emph{Journal of Fluid Mechanics}, vol. 793, pp. 41--78, 2016.

\bibitem{kramer2016tangential}
B.~Kramer and S.~Gugercin, ``Tangential interpolation-based eigensystem
  realization algorithm for mimo systems,'' \emph{Mathematical and Computer
  Modelling of Dynamical Systems}, vol.~22, no.~4, pp. 282--306, 2016.

\bibitem{li2011noise}
P.~Li, S.~Hu, and H.~Li, ``Noise issues of modal identification using
  eigensystem realization algorithm,'' \emph{Procedia engineering}, vol.~14,
  pp. 1681--1689, 2011.

\bibitem{zeiger1974approximate}
H.~p. Zeiger and A.~McEwen, ``Approximate linear realizations of given
  dimension via ho's algorithm,'' \emph{IEEE Transactions on Automatic
  Control}, vol.~19, no.~2, pp. 153--153, 1974.

\bibitem{kutz2013data}
J.~N. Kutz, \emph{Data-driven modeling \& scientific computation: methods for
  complex systems \& big data}.\hskip 1em plus 0.5em minus 0.4em\relax Oxford
  University Press, 2013.

\bibitem{s2d}
P.~J. Schmid, ``Dynamic mode decomposition of numerical and experimental
  data,'' \emph{Journal of fluid mechanics}, vol. 656, pp. 5--28, 2010.

\bibitem{9038561}
M.~N. {Murshed} and M.~{Monir Uddin}, ``Time delay coordinate based dynamic
  mode decomposition of a compressible signal,'' in \emph{2019 22nd
  International Conference on Computer and Information Technology (ICCIT)},
  2019, pp. 1--5.

\bibitem{brunton2014state}
S.~L. Brunton, S.~T. Dawson, and C.~W. Rowley, ``State-space model
  identification and feedback control of unsteady aerodynamic forces,''
  \emph{Journal of Fluids and Structures}, vol.~50, pp. 253--270, 2014.

\bibitem{chahlaoui2002collection}
Y.~Chahlaoui and P.~Van~Dooren, ``A collection of benchmark examples for model
  reduction of linear time invariant dynamical systems.'' 2002.

\bibitem{tu2012improved}
J.~H. Tu and C.~W. Rowley, ``An improved algorithm for balanced pod through an
  analytic treatment of impulse response tails,'' \emph{Journal of
  Computational Physics}, vol. 231, no.~16, pp. 5317--5333, 2012.

\end{thebibliography}
\bibliographystyle{IEEEtran} 
\section*{Appendix}
\subsection*{Derivation of the reduced system matrices from ERA}
Let's consider a possible scenario where the Hankel matrix and the time shifted Hankel matrix are defined as,
\[H=\begin{bmatrix}
    y_{1} & y_{2} & y_{3} \\
    y_{2} & y_{3} & y_{4} \\
    y_{3} & y_{4} &  y_{5} \\
    \end{bmatrix}=\begin{bmatrix}
    CB & CAB & CA^{2}B \\
    CAB & CA^{2}B & CA^{3}B \\
    CA^{2}B & CA^{3}B &  CA^{4}B \\
    \end{bmatrix}= \bar{O}\bar{C}
    \]
\[H'=\begin{bmatrix}
    y_{2} & y_{3} &  y_{4} \\
    y_{3} & y_{4} & y_{5} \\
    y_{4} & y_{5} &  y_{6} \\
    \end{bmatrix}=\begin{bmatrix}
    CAB & CA^{2}B & CA^{3}B \\
    CA^{2}B & CA^{3}B & CA^{4}B \\
    CA^{3}B & CA^{4}B &  CA^{5}B \\
    \end{bmatrix}=\bar{O}A\bar{C}.
    \]
    It is important to note that the Hankel matrices can also be written in terms of the observability and controllability. Controllability refers to how the inputs can excite the states and observability means how the states can affect the outputs. The singular value decomposition of the Hankel matrix reads, 
$$ H = U\Sigma V^{T}$$
$$ H = U T^{2} V^{T} $$
where \(\Sigma =T^{2}\).
\(T\) will then be used to define the observability and controllability as, 
$$ H = \bar{O}\bar{C} = U T^{2} V^{T}$$ 
$$ \rightarrow \bar{O} = U T, \ \bar{C}=T V^{T}. $$
Finally, we construct the reduced system matrices 
$$ H' = \bar{O}A\bar{C} $$ 
$$ H' = U T A T V^{T}$$
$$ \rightarrow \hat{A} = T^{-1}U^{T} H' V T^{-1}.$$

\subsection*{Descriptor system for the SLICOT based problems}
The last two examples in this paper used systems from \cite{chahlaoui2002collection} which is a collection of benchmark problems that have real-life applications. The collection essentially gives the matrices (\(A, B, C, D, E\)) for  different dense, sparse and second order state space models. They are based on the following descriptor system,
\begin{equation}
E\textbf{x}_{i+1} = A\textbf{x}_{i} + B \textbf{u}_{i}
\end{equation}
\begin{equation}
\textbf{y}_{i} = C\textbf{x}_{i} + D\textbf{u}_{i},
\end{equation}
where \(E\) is invertible. Additional information like \textbf{Hankel singular values, frequency} and \textbf{frequency response} are also available in these files.

\end{document}